\documentclass[12pt]{article}
\usepackage[dvips]{color}
\usepackage{epsfig}
\usepackage{amsmath}
\usepackage{graphicx}
\def\Box{\hbox{$\rlap{$\sqcup$}\sqcap$}}
\begin{document}
\title{\bf Non-minimally Coupled Quintom Model Inspired by String Theory}
\author{J. Sadeghi $^{a}$\thanks{Email: pouriya@ipm.ir}\hspace{1mm}
M. R. Setare $^{b}$\thanks{Email:rezakord@ipm.ir }\hspace{1mm},
 A. Banijamali $^{a}$\thanks{Email: abanijamali@umz.ac.ir}\hspace{1mm} and F. Milani $^{a}$  \\
$^a$ {\small {\em  Sciences Faculty, Department of Physics, Mazandaran University,}}\\
{\small {\em P .O .Box 47415-416, Babolsar, Iran}}\\
$^b$ {\small {\em Department of Science, Payame Noor University,
Bijar, Iran}}}

\maketitle

\begin{abstract}
\noindent In this paper we consider a quintom model of dark energy
with a single scalar field $T$ given by a Lagrangian which inspired
by tachyonic Lagrangian in string theory. We consider non-minimal
coupling of tachyon field to the scalar curvature, then we obtain
the equation of state (EoS), and the condition required for the
model parameters when $\omega$ crosses over $-1$.
\\

{\bf Keywords:}Quintom model; Tachyon; Non-minimal coupling.

\end{abstract}
\newpage
\section{Introduction}
Many cosmological observations, such as SNe Ia [1], WMAP [2], SDSS
[3], Chandra X-ray observatory [4] etc., discover that our universe
is undergoing an accelerated expansion. They also suggest that our
universe is spatially flat, and consists of about
$70~^{\circ}/_\circ$ dark energy (DE) with negative pressure,
$30~^{\circ}/_\circ$ dust matter (cold dark matter plus baryons),
and negligible radiation. Dark energy has been one of the
most active fields in modern cosmology [5].\\
In modern cosmology of dark energy, the equation of state parameter
(EoS) $\omega=\frac{p}{\rho}$ plays an important role, where $p$ and
$\rho$ are its pressure and energy density, respectively. To
accelerate the expansion, the EoS of dark energy must satisfy
$\omega<-\frac{1}{3}$. The simplest candidate of the dark energy is
a tiny positive time-independent cosmological constant $\Lambda$,
for which $\omega=-1$. However, it is difficult to understand why
the cosmological constant is about 120 orders of magnitude smaller
than its natural expectation (the Planck energy density). This is
the so-called cosmological constant problem. Another puzzle of the
dark energy is the cosmological coincidence problem: why are we
living in an epoch in which the dark energy density and the dust
matter energy are comparable?. As a possible solution to these
problems various dynamical models of dark energy have been proposed,
such as quintessence \cite{{c6},{c7}}. The analysis of the
properties of dark energy from recent observations mildly favor
models with $\omega$ crossing -1 in the near past. So far, a large
class of scalar-field dark energy models have been studied,
including tachyon \cite{tachyon}, ghost condensate
\cite{ghost1,ghost2} and quintom \cite{{c9},{c12},{c14},{c15}}, and
so forth.  In addition, other proposals on dark energy include
interacting dark energy models \cite{intde}, braneworld models
\cite{brane}, and holographic dark energy modeles \cite{holo}, etc.
The Ref.\cite{c9}  is the first paper showing explicitly the
difficulty of realizing $\omega$ crossing over -1 in the
quintessence and phantom like models. Because it has been proved
\cite{{c9},{c10},{c11}} that the dark energy perturbation would be
divergent as the equation of state $\omega$ approaches to -1. The
quintom scenario of dark energy is designed to understand the nature
of dark energy with $\omega$ across -1. The quintom models of dark
energy differ from the quintessence, phantom and k-essence and so on
in the determination of the cosmological evolution and the fate of
the universe.
\\To realize a viable quintom scenario of dark energy it
needs to introduce extra degree of freedom to the conventional
theory with a single fluid or a single scalar field. The first model
of quintom scenario of dark energy is given by Ref.\cite{c9} with
two scalar fields. This model has been studied in detail later on
\cite{{c12},{c14},{c15}} (to see the bouncing solution in the
universe dominated by quintom matter refer to \cite{bou}). Recently
there has been an upsurge in activity for constructing such model in
string theory \cite{c16}. In the context of string theory, the
tachyon field in the world volume theory of the open string
stretched between a D-brane and an anti-D-brane or a non-BPS D-brane
plays the role of scalar field in the quintom model \cite{c17}. The
effective action used in the study of tachyon cosmology consists of
the standard Einstein-Hilbert action and an effective action for the
tachyon field on unstable D-brane or D-brane anti D-brane system.
What distinguishes the tachyon action from the standard Klein-
Gordon form for scalar field is that the tachyon action is
non-standard and is of the " Dirac-Born-Infeld " form \cite{c18}.
The tachyon potential is derived from string theory itself and has
to satisfy some definite properties to describe
tachyon condensation and other requirements in string theory.\\
In this paper, we consider an action for tachyon non-minimally
coupled to gravity \cite{c19} inspired by the string theory. An
outline of this paper is as follows. In section 2 we introduce
action for tachyon non-minimally coupled to gravity with an extra
term. By performing a conformal transformation we obtain the new
action. In order to discuss the equation of state we derive the
corresponding energy density and pressure for this model. By solving
this equation we obtain the conditions required for the $\omega$
across -1. Section 3 is devoted to discussion of our results.

\section{Non-minimally coupled tachyon gravity with extra term}
We consider the action Ref.\cite{c20} for tachyon non-minimally
coupled to gravity, in this action we add an extra term $T\Box {T}$
to the usual terms in the square root. In that case the following
action is the same as Ref.\cite{c21} just different to the $ Rf(T),$
\begin{equation}
S=\int d^{4}x \sqrt{-g} \left[\frac{M_{P}^{2}}{2}Rf(T) -
AV(T)\sqrt{1-\alpha'g^{\mu\nu}\partial_{\mu}T\partial_{\nu}T+\beta'T
\Box T}\right],
\end{equation}
where $f(T)$ is a function of the tachyon $T$ and corresponds to the
non-minimal coupling factor. Here $V(T)$ is the tachyon potential
which is bounded and reaching its minimum asymptotically.
$M_{P}=\frac{1}{\sqrt{8\pi G}}$ is reduced
Planck mass.\\
Action (1) generalizes the usual "Born- Infeld- type" action for the
effective description of tachyon dynamics which can be obtained by
the stringy computations for a non- BPS D3- brane in type II theory.
The extra term in action (1) has a significant cosmological
consequence, so we cannot exclude its existence in an action such as
the "Born- Infeld- type" action.\\
The model with operator $T \Box T$ for realizing of $\omega$
crossing -1 has been proposed in \cite{c12}. They considered a
dimension-6 operator as $ (\Box T)^{2}$. However in the present
paper, the operator $T \Box T$ appears at the same order as the
operator $\partial_{\mu}T\partial^{\mu}T$ does in the "Born- Infeld-
type"
action and also we take into account scalar curvature non-minimally coupled to the tachyon field.\\
The action (1) can be brought to the simpler form to derive the
equation of motion, energy density and pressure, by performing a
conformal transformation as follows:
\begin{eqnarray}
g_{\mu\nu}\longrightarrow f(T) g_{\mu\nu}.
\end{eqnarray}
The above conformal transformation yields to the following action:\\
$$S=\int d^{4}x\sqrt{-g}\
\Bigg[\frac{M_{P}^{2}}{2}(R-\frac{3}{2}\frac{f'^{2}}{f^{2}}\partial_{\mu}T\partial^{\mu}T)$$

\begin{eqnarray}
-A\tilde{V}(T)\sqrt{1-(\alpha'f(T)-2\beta'f'(T)T)\partial_{\mu}T\partial^{\mu}T+\beta'f(T)T
\Box T}\Bigg]\,
\end{eqnarray}
where $\tilde{V}(T)=\frac{V(T)}{f^{2}}$ is now the effective
potential of the tachyon.\\
 For a flat Friedman- Robertson- Walker
(FRW) universe and a homogenous scalar field $T$, the equation of
motion can be solved equivalently by the following two equations,
$$\ddot{\psi}+3H\dot{\psi}=(\frac{2\beta'f'T-\alpha'f}{fT})\dot{\psi}\dot{T}-\frac{A^{2}\beta'f\tilde{V}T}{2\psi}\tilde{V}
'-\frac{3M_{P}^{2}}{2}(\frac{ff'f''-f'^{3}}{f^{3}})\dot{T}^{2}$$
\begin{eqnarray}
-\left[\frac{(1-\beta')(\alpha'-2\beta')}{\beta'}\frac{f'}{f}-\frac{\alpha'}{T}
\right]\frac{\psi\dot{T}^{2}}{T},
\end{eqnarray}
\begin{eqnarray}
\ddot{T}+3H\dot{T}=\frac{2\left[(\frac{ff''+\beta'f'}{f^{2}})T\dot{T}^{2}-2
(\alpha'-2\beta'\frac{f'}{f}T)H\dot{T}\right]}{1+\frac{2\alpha'}{\beta'}-3\frac{f'}{f}T-
\frac{3M_{P}^{2}}{2}(\frac{f'}{f})^{2}\frac{T}{\psi}},
\end{eqnarray}

where $$\psi=\frac{\partial  \mathcal{L}}{\partial \Box
T}=-\frac{A\beta'\tilde{V}fT}{2h}$$ $$
h=\sqrt{1-(\alpha'f-2\beta'f'T)\partial_{\mu}T\partial^{\mu}T+\beta'fT
\Box T}$$ and $$\tilde{V}^{'}=\frac{d\tilde{V}}{dT}.$$
$H=\frac{\dot{a}}{a}$ is
the Hubble parameter.\\
The energy momentum tensor $T^{\mu\nu}$ is given by the standard
definition: $$\delta_{g_{\mu\nu}}S=-\int
d^{4}x\frac{\sqrt{-g}}{2}T^{\mu\nu}\delta g_{\mu\nu}.$$ So the
energy density, pressure and Friedman equation are found to be
\begin{eqnarray}
\rho=A\tilde{V}h+\frac{d}{a^{3}dt}(a^{3}\psi\dot{T})+(\alpha'f-2\beta'f'T)
\frac{A\tilde{V}}{h}\dot{T}^{2}-2\dot{\psi}\dot{T}+\frac{3M_{P}^{2}}{4}(\frac{f'}{f}^{2})\dot{T}^{2},
\end{eqnarray}
\begin{eqnarray}
p=-A\tilde{V}h-\frac{d}{a^{3}dt}(a^{3}\psi\dot{T})+\frac{3M_{P}^{2}}{4}(\frac{f'}{f}^{2})\dot{T}^{2},
\end{eqnarray}

\begin{eqnarray}
H^{2}=\frac{A}{3M_{P}^{2}}\tilde{V}h+\frac{d}{3M_{P}^{2}a^{3}dt}(a^{3}\psi\dot{T})+\frac{(\alpha'f-2\beta'f'T)}
{3M_{P}^{2}}\frac{A\tilde{V}\dot{T}^{2}}{h}
-\frac{2}{3M_{P}^{2}}\dot{\psi}\dot{T}+\frac{1}{4}(\frac{f'}{f}^{2})\dot{T}^{2}.
\end{eqnarray}
 We now study the cosmological evolution of equation of state for
the present model. The equation of state is $p=\omega \rho$. To
explore the possibility of the $\omega$ across -1, we have to check
$\frac{d}{dt}(\rho+p)\neq0$ when $\omega\longrightarrow-1$. \\
From equations (6) and (7) one can obtain the following expressions,
\begin{eqnarray}
\rho+p=\frac{3M_{P}^{2}}{2}(\frac{f'}{f}^{2})\dot{T}^{2}+(\alpha'f-2\beta'f'T)
\frac{\tilde{V}}{h}\dot{T}^{2}+A\beta'\frac{d}{dt}(\frac{f\tilde{V}T}{h})\dot{T},
\end{eqnarray}
and
\begin{eqnarray}
\frac{d}{dt}(\rho+p)&=&\left[3M_{P}^{2}(\frac{f'}{f}^{2})\dot{T}+A\beta'\frac{d}{dt}
(\frac{f\tilde{V}T}{h})\right]\ddot{T}+
\left[(\alpha'\dot{f}-2\beta'\dot{f}')
\frac{\tilde{V}}{h}+3M_{P}^{2}\frac{f'}{f}\frac{d}{dt}(\frac{f'}{f})\right]\dot{T}^{2}
\nonumber\\&-&2\beta'f'\frac{\tilde{V}}{h}\dot{T}^{3}+(\alpha'f-2\beta'f'T)\frac{d}{dt}(\frac{\tilde{V}\dot{T}^{2}}{h})+A\beta'\frac{d^{2}}
{dt^{2}}(\frac{f\tilde{V}T}{h})\dot{T}.
\end{eqnarray}
From equation (9) one finds  $\dot{T}=0$,
$\hspace{.25cm}$or$\hspace{.25cm}$
$\left[\frac{3M_{P}^{2}}{2}(\frac{f'^{2}}{f})+(\alpha'f-2\beta'f'T)\frac{\tilde{V}}{h}\right]\dot{T}=
-A\beta'\frac{d}{dt}(\frac{f\tilde{V}T}{h})$ when
$\omega\longrightarrow-1$.\\
 If $\dot{T}=0$ then equation (10) gives,
\begin{eqnarray}
\frac{d}{dt}(\rho+p)=A\beta'\frac{d}{dt}(\frac{f\tilde{V}T}{h})\ddot{T}=-\frac{\tilde{V}f^{2}
}{2h^{3}}\beta'^{2}T^{2}\ddot{T}\frac{d}{dt}\Box T.
\end{eqnarray}
So we need $T\neq0$, $\ddot{T}\neq0$ and $\frac{d}{dt}\Box T\neq0$,
for having $\omega$ across -1. With these crossing over condition
and asymptotically behavior of $V(T)$ one concludes that crossing
over -1 must happen before reaching the potential minimum
asymptotically. This implies that when crossing over -1 occur the
field $T$ must continue to run away as it should be since we have
$\ddot{T}\neq0$. One can obtain the same result with the second
condition and also note that in the second case $\dot{T}$ is
non-zero.\\
Now we consider two specific examples to see how EoS evolves in our
model. In numerical calculations we have used the expansion of
$V(T)=e^{-\lambda T^{2}}$ (motivated from boundary string field
theory \cite{c21}) in Figure 1 and $V(T)=\frac{V_{0}}{e^{\lambda
T}+e^{-\lambda
T}}$ in Figure 2, also we take $f(T)=1+\sum_{i=1}c_{i}T^{2i}.$\\

\begin{tabular*}{2cm}{cc}
\hspace{0.25cm}\includegraphics[scale=0.25]{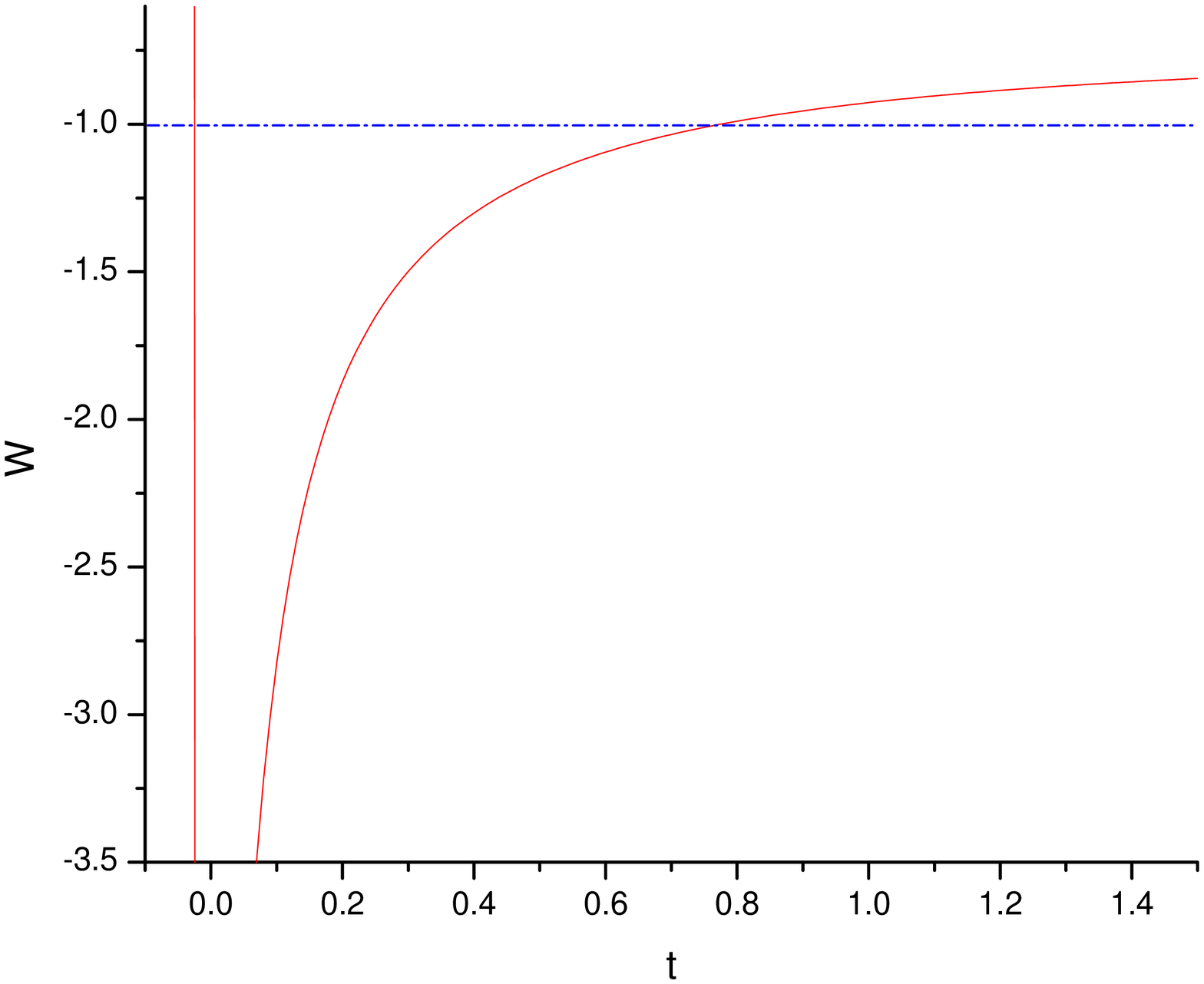}\hspace{0.5cm}\\
\hspace{2
cm}\textbf{Figure 1:} \,The plot of EoS  for the potential
$V(T)=e^{-\lambda T^{2}}$,\\~~~~~~~~~~
 $\alpha=-2$ and $\beta=2.2$. Initial values are $\phi=1$ and $\dot{\phi}=3$. \\
\end{tabular*}\\\\\\
\begin{tabular*}{2cm}{cc}
\hspace{0.25cm}\includegraphics[scale=0.25]{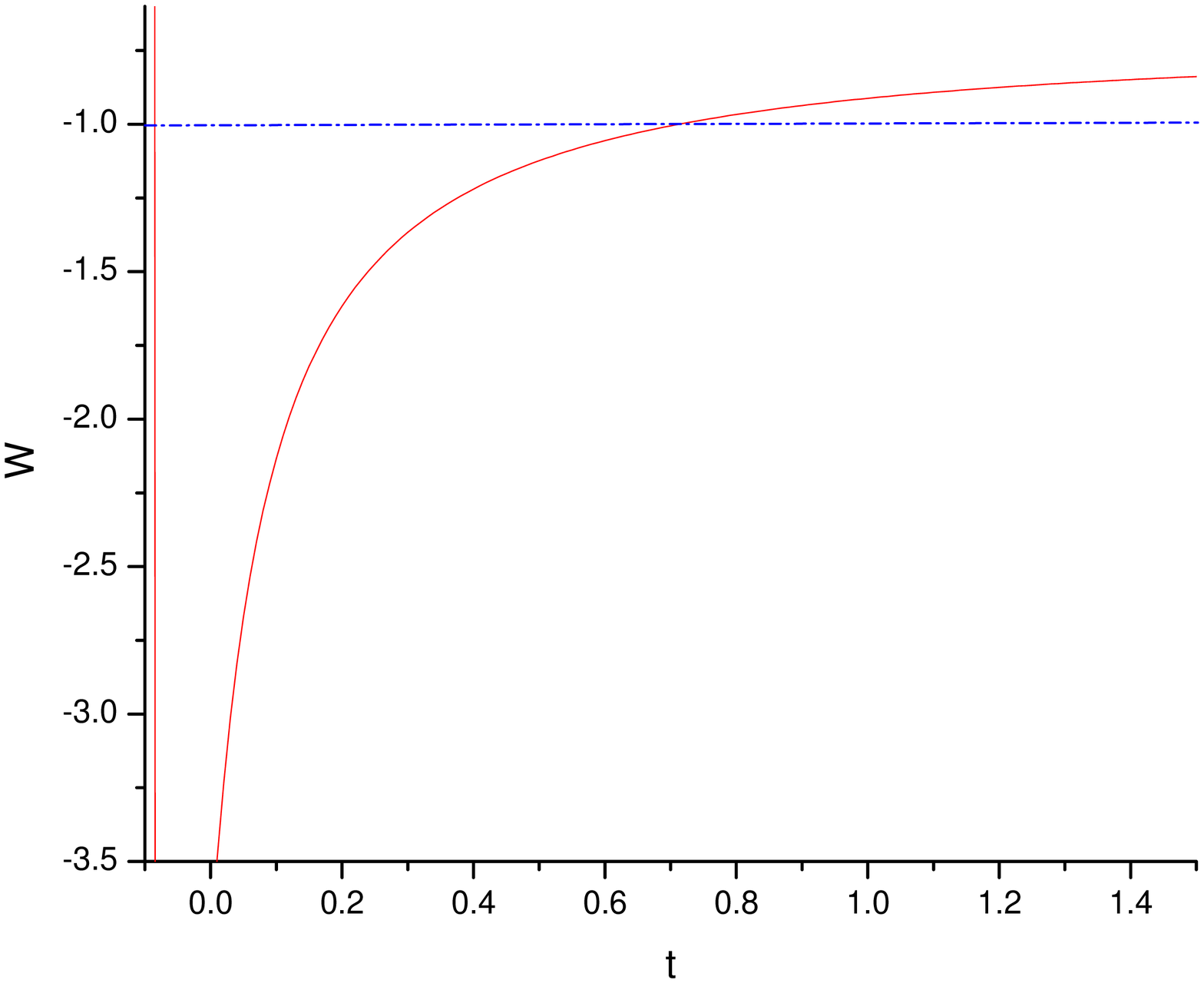}\hspace{0.5cm}\\
\hspace{1.7cm}\textbf{Figure 2:} \,The plot of EoS  for the
potential $V(T)=\frac{V_{0}}{e^{\lambda T}+e^{-\lambda
T}}$,\\~~~~~~~~~~~~~$\alpha=-2$, $\beta=2.2$, $\lambda=2$ and $V_{0}=5$. Initial values are $\phi=1$ and $\dot{\phi}=-3$. \\
\\
\end{tabular*}\\\\\\
\section{Conclusion}

In order to solve cosmological problems and because the lack of our
knowledge, for instance to determine what could be the best
candidate for DE to explain the accelerated expansion of universe,
the cosmologists try to approach to best results as precise as they
can by considering all the possibilities they have. Within the
different candidates to play the role of the dark energy, the
quintom model, has emerged as a possible model  with EoS across
$-1$. In this paper we have introduced a string inspired quintom
model non-minimally coupled to gravity with an extra term in the
usual effective action of tachyon dynamics. This modification has
been done due to crossing over -1 of EoS. We showed that crossing
over -1 occur in this model, before reaching the tachyon potential
asymptotically to its minimum.
 
\end{document}